
\magnification=1200
\baselineskip=20pt
\overfullrule=0pt
\def\bull{\vrule height .9ex width .8ex depth}
\def\bull{\vrule
height .9ex width .8ex depth}
\centerline{\bf PARA-STATISTICS AS LIE-SUPER TRIPLE SYSTEMS}
\bigskip
\centerline{by}
\bigskip
\centerline{Susumu Okubo}
\centerline{Department of Physics and Astronomy}
\centerline{University of Rochester}
\centerline{Rochester, NY 14627}

\vskip 1.5in

\noindent \underbar{Abstract}

We first reformulate para-statistics in terms of Lie-super triple systems.
In this way, we reproduce various new kinds of para-statistics discovered
recently by Palev in addition to the standard one.  Also, bosonic and
fermionic operators may not necessarily commute with each other.

\vskip 2in


\vfill\eject

\noindent {\bf 1. \underbar{Introduction}:}

The para-statistics$^{1)}$ is a generalization of familiar bosonic and
fermionic statistics.  As an example, consider the case of the para-fermi
statistics, where annihilation and creation operators respectively given by
 $a_j$ and $a^\dagger_j$ $(j = 1,2,\dots,n)$ satisfy triple commuation
relations
$$\eqalignno{[[a_j, a^\dagger_k ], a_\ell] &= 2 \delta_{k \ell} a_j \quad ,
&(1.1a)\cr
[[a_j , a^\dagger_k ], a^\dagger_\ell ] &= -2 \delta_{j \ell} a^\dagger_k
\quad , &(1.1b)\cr
[[a_j , a_k], a^\dagger_\ell] &= \delta_{k \ell} a_j -
 \delta_{j \ell} a_k \quad , &(1.1c)\cr
[[a^\dagger_j , a^\dagger_k], a_\ell] &= \delta_{k \ell} a^\dagger_j
- \delta_{j \ell} a^\dagger_k \quad , &(1.1d)\cr
[[a_j , a_k ],a_\ell] &= [[a^\dagger_j , a^\dagger_k ],
a^\dagger_\ell ] = 0 \quad . &(1.1e)\cr}$$
It is well known that these equations are equivalent to the Lie algebra
 so(2n+1), while the corresponding case of the para-boson leads$^{2)}$ to
the Lie-super algebra$^{3)}$ osp(1$|$2n).

More recently, Palev$^{4),5)}$ has found several new para-bosonic
statistics with different Lie-super algebra structures such as
 s$\ell$(1$|$n) and
osp(3$|$2) etc.  Especially, the 3-dimensional bosonic harmonic oscillator
model with the osp(3$|$2) structure gives the spin 1/2 realization.

The purpose of this paper is twofold.  We first reformulate the
para-statistics as Lie-super triple system which is a generalization of the
standard Lie-triple system$^{6),7)}$.  One advantage of the new formulation
is that the underlying Lie-super algebraic structure of the para-statistics
is manifest.  Second, we can construct various new kinds of
para-statistics, including those by Palev$^{4),5)}$ in a natural way by
this method.  Especially, we find para-systems where boson and fermions do
not necessarily commute with each other.

We organize this paper as follows.  In section 2, we will briefly sketch
the notion of Lie-super triple systems and explain their relationship with
Lie-super algebras.  In section 3, we will give various examples of
para-statistics of new kinds.  In Appendix A, some explicit constructions
of Lie-super algebras in terms of Lie-super triple systems will be given,
and in Appendix B, the relationship between Lie-super triple systems and
ortho-symplectic super-triple systems will be explained.

\medskip

\noindent {\bf 2. \underbar{Lie-super Triple System}:}

Let $V$ be a vector space over a field $F$ which is assumed to be either
real or complex.  If $V$ is a linear span of elements $x_1 , x_2 , \dots ,
x_n$, then we write
$$V = \ < x_1 , x_2 , \dots , x_n> \quad . \eqno(2.1)$$
We suppose that $V$ is a super-space, i.e. it is a direct sum
$$V = V_B \oplus V_F \eqno(2.2)$$
whose dimensions are specified as
$$\eqalignno{N &= N_B + N_F = \ {\rm Dim}\ V \quad , &(2.3a)\cr
N_B &= \ {\rm Dim}\ V_B \quad , \quad N_F = \ {\rm Dim}\ V_F \quad .
&(2.3b)\cr}$$
We introduce the degree (or grade or signature) by
$$\sigma (x) = \cases{0, &if $x \ \epsilon\ V_B$\cr
\noalign{\vskip 5pt}%
1, &if $x\ \epsilon\ V_F$\cr}\quad . \eqno(2.4)$$
Any element considered in this paper is always assumed to be homogeneous,
i.e. either $x\ \epsilon\ V_B$ or $x\ \epsilon\ V_F$.

Now, the triple product $[.,.,.]$ is a tri-linear mapping
$$[.,.,.]\ ;\ V \otimes V \otimes V \rightarrow V \eqno(2.5)$$
i.e. we have

\item{(i)} $[x,y,z]\ \epsilon\ V$ for $x,\  y,\ z\ \epsilon\ V$

\item{(ii)} $[x,y,z]$ is linear in each variable $x,\ y,\ {\rm and}\ z$.
For example, we must have

$$\left[ \alpha x_1 + \beta x_2 , y, z \right] = \alpha
\left[ x_1 , y , z \right] + \beta \left[ x_2 , y , z \right]$$
for $\alpha,\ \beta\ \epsilon\ F \ {\rm and\ for}\ x_1, \ x_2,\ y,\ z \
\epsilon\ V$.
We impose further the triple-super product condition, i.e.
$$\sigma ([ x,y,z]) = \{
\sigma (x) + \sigma (y) + \sigma (z)\} \quad ({\rm mod}\ 2)
\quad . \eqno(2.6)$$
Suppose now that the triple-super product satisfies the following ans\"atz:

\item{(i)} $[x,y,z] = -(-1)^{xy} [y,x,z]$\hfill  (2.7a)

\item{(ii)} $(-1)^{xz} [x,y,z] + (-1)^{yx} [y,z,x] +
(-1)^{zy} [z,x,y] = 0$ \hfill (2.7b)

\item{(iii)} $[u,v,[x,y,z]] = [[u,v,x],y,z] + (-1)^{(u+v)x}
[x,[u,v,y],z]$
\item{     } $\qquad \qquad + (-1)^{(u+v)(x+y)} [x,y,[u,v,z]]$
\hfill (2.7c)

\noindent where we have set for simplicity
$$\eqalign{&(-1)^{xy} = (-1)^{\sigma (x) \sigma (y)}\cr
&(-1)^{(u+v)x} = (-1)^{ux +vx} = (-1)^{[\sigma (u) +
\sigma (v)]\sigma (x)}\cr}$$
etc., following the standard convention.

We call any vector space $V$ with the triple-super product $[x,y,z]$
satisfying these conditions to be a Lie-super triple system$^{8)}$.  If
$V = V_B$ with $V_F = 0$, then this notion reduces to the standard
 Lie-triple system$^{6),7)}$.  In our applications, the underlying vector
space $V$ posseses always a bilinear form $<x|y>\ \epsilon\ F$ satisfying

\item{(i)} $<x|y>\ = (-1)^{xy} <y |x>$ \hfill (2.8a)

\item{(ii)} $<x|y>\ = 0 \quad {\rm if}\quad \sigma (x) \not=
\sigma (y)$ (mod 2) \hfill (2.8b)

\noindent which we shall assume hereafter in this paper.  Moreover,
$<x|y>$ is assumed to be non-degenerate, unless it is
 stated otherwise.

Before going into further details, it may be instructive to offer some
examples of Lie-super triple systems since they are needed in the next
section.
\medskip

\noindent \underbar{Example 1}

Let $V$ be a Lie-super algebra with a
 bilinear product $[x,y]$ satisfying the following conditions

\vfill\eject

\item{(i)} $\sigma ([x,y]) = \{ \sigma (x) + \sigma (y)\}$
(mod 2) \hfill (2.9a)

\item{(ii)} $[y,x] = - (-1)^{xy} [x,y]$\hfill (2.9b)

\item{(iii)} $(-1)^{xz} [[x,y],z] + (-1)^{yx} [[y,z],x] +
(-1)^{zy} [[z,x],y] = 0 \quad .$ \hfill (2.9c)

\noindent Then, defining the triple product $[x,y,z]$ by
$$[x,y,z] = [[x,y],z] \quad , \eqno(2.10)$$
$V$ becomes a Lie-super triple systems.

\medskip

\noindent \underbar{Example 2}

Let $<x|y>$ be the bilinear form satisfying Eqs. (2.8).  For any non-zero
constant $\lambda$, we set
$$\eqalign{[x,y,z] &= \lambda \{ <y|z>x - (-1)^{xy} <x|z>y\}\cr
&= \lambda \{ <y |z>x - (-1)^{yz} <x|z>y \} \quad . \cr} \eqno(2.11)$$

\noindent Then, $[x,y,z]$ defines a Lie-super triple product, and we used
$(-1)^{xy} <x|z>\  = (-1)^{zy}$
$<x|z>$ because of Eq. (2.8b) for the 2nd
relation in Eq. (2.11).

\medskip

\noindent \underbar{Example 3}

Let $P\ :\ V \rightarrow V$ be a grade-preserving linear map in $V$, i.e.
$$\sigma (P x) = \sigma (x) \qquad {\rm for \ any}\ x\ \epsilon \  V
\quad . \eqno(2.12)$$
Moreover, we assume the validities of

\item{(i)} $P^2 = \lambda\  Id \quad (\lambda\  \epsilon\ F)$ \hfill
(2.13a)

\item{(ii)} $<x|Py>\ = - <Px |y>$ \hfill (2.13b)

\noindent where $Id$ is the identity mapping in $V$.  We now introduce the
triple product by
$$\eqalign{[x,y,z] = \ &<y|Pz>Px - (-1)^{xy} <x|Pz>Py -
2<x|Py>Pz\cr
&+ \lambda <y|z>x - \lambda (-1)^{xy} <x|z>y \quad .\cr} \eqno(2.14)$$
After some calculations, we can verify that $V$ becomes a Lie-super triple
system with this triple product.

\medskip

\vfill\eject

\noindent \underbar{Example 4}

Let $E_\alpha\  (\alpha = 1,2,3)$ be grade-preserving
linear mappings in $V$,
i.e.
$$\sigma \left( E_\alpha x \right) = \sigma (x) \quad , \quad
(\alpha = 1,2,3) \eqno(2.15)$$
satisfying the quaternionic relation
$$E_\alpha E_\beta = \lambda \delta_{\alpha \beta}
Id + c \sum^3_{\gamma=1} \epsilon_{\alpha \beta \gamma} E_\gamma
\eqno(2.16a)$$
for some non-zero constants $\lambda$ and $c$, as well as
$$<x|E_\alpha y>\ = -<E_\alpha x| y> \quad . \eqno(2.16b)$$
Here $\epsilon_{\alpha \beta \gamma}$ is the totally antisymmetric Levi-Civita
symbol in the 3-dimension.  Now, we set
$$\eqalign{[x,y,z] = &\sum^3_{\alpha =1} <y|E_\alpha z> E_\alpha x -
(-1)^{xy} \sum^3_{\alpha = 1} <x |E_\alpha z> E_\alpha y\cr
&-2 \sum^3_{\alpha = 1} <x |E_\alpha y> E_\alpha z +
\lambda <y|z>x - \lambda (-1)^{xy}
<x|z>y \cr}\eqno(2.17)$$
which can be shown to satisfy all axioms of the Lie-super triple products.

\medskip

\noindent \underbar{Example 5}

Let $P_\mu$ for any integer $\mu$ be grade-preserving mappings in $V$,
satisfying the condition
$$<x|P_\mu y>\ = \ <P_\mu x|y> \quad . \eqno(2.18)$$
Note the change of sign in Eq. (2.18) in comparison to that of Eq.
(2.16b).  Moreover, we assume the validities of

\item{(i)} $P_{\mu +n} = P_\mu$ for any integer $\mu$ \hfill
(2.19a)

\item{(ii)} $P_n = P_0 = Id$ \hfill (2.19b)

\item{(iii)} $P_\mu P_\nu = P_{\mu + \nu}$ \hfill (2.19c)

\noindent for a positive integer $n$.  Especially, $P_\mu$'s are periodic
with the period $n$.  Then,
$$[x,y,z] = \sum^n_{\mu =1} \left\{ P_{n-\mu} x< y| P_\mu z> \ -
\ (-1)^{xy} P_{n - \mu} y< x|P_\mu z>\right\}
\eqno(2.20)$$
defines a Lie-super triple product.

\medskip

\noindent \underbar{Remark 1}

The examples 2, 3, and 4 given above are intimately related to the
ortho-symplectic super-triple system$^{8)}$.  See Appendix B.  Moreover,
there exists another way of constructing a Lie triple system from a class
of triple systems called Freudenthal-Kantor $U(\varepsilon)$
systems$^{9),10)}$.  This fact may be generalizable for the case of
super-systems as in the case of ortho-symplectic super-triple
system$^{11)}$. $\bull$

Returning to the general case, we can construct a Lie-super algebra from
any Lie-super triple system as follows.  Our method is a straightforward
generalization of the so-called canonical construction$^{6),7)}$ of Lie
algebras out of Lie triple systems.  Let $V$ be a Lie-super triple system.
For any $x,\ y\ \epsilon\ V$, we introduce the left-multiplication operator
$L_{x,y} \ :\ V \rightarrow V$ by
$$L_{x,y} z = [ x,y,z] \eqno(2.21)$$
We also write for some time $L_{x,y}$ as
$$L(x,y) = L_{x,y} \eqno(2.22)$$
whenever it is more convenient to do so.  We note that a set of all
 $L_{x,y} (x,y\ \epsilon\ V)$ generates an \underbar{associative} enveloping
algebra.  Defining the super commutator in the enveloping algebra by
$$\left[ L_{u,v},L_{x,y} \right] = L_{u,v} L_{x,y} -
(-1)^{(x+y)(u+v)} L_{x,y} L_{u,v} \quad , \eqno(2.23)$$
we see then that Eqs. (2.7a) and (2.7c) are rewritten as

\item{(i)} $L_{y,x} = - (-1)^{xy} L_{x,y}$ \hfill (2.24a)

\item{(ii)} $[L_{u,v} , L_{x,y}] =
L_{[u,v,x],y} + (-1)^{(u+v)x} L_{x,[u,v,y]} \quad .$
\hfill (2.24b)

\noindent Defining the grade of $L_{x,y}$ now by
$$\sigma \left( L_{x,y}\right) = (\sigma (x) + \sigma (y)) \quad
({\rm mod}\ 2) \quad , \eqno(2.25)$$
Eqs. (2.24) imply that they define a Lie-super algebra when we note
$$\left[ L_{u,v} , L_{x,y} \right] = - (-1)^{(u+v)(x+y)}
\left[ L_{x,y} , L_{u,v} \right] \quad . \eqno(2.26)$$
To be more definite, let
$$\eqalign{M &= \ <L_{x,y} \ ;\ x,\ y\ \epsilon\ V>\cr
&= \ {\rm linear\ vector\ space\ spanned\ by\ all}\
L_{x,y}\ (x,\ y\ \epsilon\ V) \quad .\cr} \eqno(2.27)$$
Then, $M$ is a Lie-super algebra.

Actually, we can construct a larger Lie-super algebra $L$
as follows.  Consider a vector space direct sum given by
$$L = V \oplus M \eqno(2.28)$$
and define a bilinear product in $L$ by
$$\eqalignno{&[x,y] = L_{x,y} &(2.29a)\cr
&[L_{x,y} , z ] = -(-1)^{(x+y)z} [z, L_{x,y} ] =
[x,y,z] &(2.29b)\cr}$$
for $x,\ y,\ z \ \epsilon\ V$ together with Eq. (2.23) for
$[M,M]$.  Then, it is not hard to see that $L$ defines a Lie-super algebra.
 For example, we calculate
$$\eqalign{(-1)^{xz} &[[x,y],z] + (-1)^{yx} [[y,z],x]
+ (-1)^{zy} [[z,x],y]\cr
&= (-1)^{xz} [L_{x,y},z] + (-1)^{yx} [L_{y,z},x]
+ (-1)^{zy} [L_{z,x},y]\cr
&= (-1)^{xz} [x,y,z] + (-1)^{yx} [y,z,x] + (-1)^{zy}
[z,x,y]\cr
&= 0\cr}$$
by Eqs. (2.29) and (2.7b).  If $V = V_B$ with $V_F = 0$, this procedure
reproduces, of course, the canonical construction$^{6),7)}$ of a Lie
algebra out of a Lie triple system.

We see  from Eqs. (2.24) and (2.29) that
$$\eqalignno{[M,M] &\subseteq M \quad , &(2.30a)\cr
[M,V] &\subseteq V \quad , &(2.30b)\cr
[V,V] &\subseteq M \quad , &(2.30c)\cr}$$
which is familiar in the theory of symmetric homogeneous space$^{12)}$.
Therefore, if $G$ and $H$ are Lie-super groups generated respectively by
$L$ and $M$, we may write
$$V = G/H \quad . \eqno(2.31)$$
Moreover, Eqs. (2.29) imply the validity of
$$[x,y,z] = [[x,y],z] \quad . \eqno(2.32)$$
Although the relation Eq. (2.32) is formally identical in form to Eq.
(2.10) of the example 1, their meanings are entirely different.  For the
present case, we have $[x,y]\ \epsilon \ M$, while the commutator $[x,y]$
in Eq. (2.10) is an element of $L = V$.  Strictly speaking, $[x,y]$ for the
present case is not well-defined in $V$ but only for a larger space $L$,
although the triple product $[x,y,z]$ is well-defined in $V$.  The
situation is essentially the same$^{12)}$
 for the case of the symmetric homogeneous
spaces.

\medskip

\noindent \underbar{Remark 2}

\indent From the definition Eq. (2.27) we see
$${\rm Dim}\ M \leq {1 \over 2}\ N_B \left( N_B + 1 \right) +
{1 \over 2}\ N_F \left( N_F - 1 \right) +
N_B N_F \quad . \eqno(2.33)$$
However, the equality in Eq. (2.33) does not hold in general.  This is due
to the fact that $L_{x,y}$'s may not necessarily be linearly independent of
each other.  Interchanging $x \leftrightarrow u$ and $y \leftrightarrow v$
in Eq. (2.24b) and noting Eq. (2.26), we find
$$\eqalign{\big[ L_{u,v},L_{x,y}\big] &= L_{[u,v,x],y} +
(-1)^{(u+v)x} L_{x,[u,v,y]}\cr
&= -(-1)^{(u+v)(x+y)} L_{[x,y,u],v} -
(-1)^{(u+v)y} L_{u,[x,y,v]}\cr}\eqno(2.34)$$
which imposes some linear relations among $L_{x,y}$'s.
\medskip

\noindent {\bf 3. \underbar{Para-statistics}:}
\medskip

\noindent {\bf 3.1 \underbar{Example 1}}

We base our construction of para-statistics upon the validity of
Eq. (2.32).  Before
 going into detail, we first remark that we may forsake the identification
of the space $M$ with the linear span of $L_{x,y}$'s as in Eq.
(2.27), by regarding Eq. (2.28)  as an
explicit realization of the Lie-super algebra $L$ satsifying conditions of
Eqs. (2.30).  Moreover, we may discuss the example 1 also by assuming $M
\subseteq V$ with $L = V$, if we wish.
We will illustrate our method in what follows below.

\medskip

\noindent {\bf 3.2 \underbar{Example 2}}

First let us consider the simplest example given by Eq. (2.11).  As we will
show in the Appendix, Lie-super algebras $M$ and $L$ are specified to be
$$\eqalignno{M &= \ {\rm osp} (N_B | N_F) \quad , &(3.1a)\cr
L &= \ {\rm osp}(N_B + 1 | N_F) \quad , &(3.1b)\cr}$$
provided that the bilinear form $<x|y>$ is non-degenerate.
Here osp stands$^{3)}$ for the ortho-symplectic Lie-super algebra.
 Suppose first
the case of $N_F = 0$ i.e. $V_F = 0$, and identify
$$\eqalignno{&V = V_B = \ <a_j, a^\dagger_j, j=1,2,\dots,n> &(3.2a)\cr
&<y|x> \ =\ <x|y> &(3.2b)\cr}$$
together with
$$\eqalignno{&\sigma (a_j) = \sigma \big( a^\dagger_j \big) = 0
\quad , &(3.3a)\cr
&< a_j |a^\dagger_k>\  = \ <a^\dagger_k | a_j>\ = \delta_{jk} \quad ,
&(3.3b)\cr
&< a_j |a^\dagger_k>\  = \ <a^\dagger_j | a^\dagger_k>\ = 0 \quad .
&(3.3c)\cr}$$
Choosing $\lambda = 2$, Eqs. (2.11) and (2.32) give
$$[[x,y],z] = 2<y|z>x - 2<x|z>y \quad .
\eqno(3.4)$$
Especially, identifying $x= a_j,\ y = a^\dagger_k,\
{\rm and}\ z = a_\ell\ ({\rm or}\ a^\dagger_\ell)$, we find Eqs. (1.1a) and
(1.1b).  Similarly, the rest of equations (1.1c), (1.1d), and (1.1e)
follows from Eq. (3.4) for suitable choices of $x,\ y,\ z\ \epsilon\ V$.
Therefore, the present example reproduces the para-fermi statistics with
$$M = \ {\rm osp} (2n|0) = \ {\rm so} (2n) \qquad {\rm and}\qquad
L =\ {\rm osp} (2n +1|0) = \ {\rm so} (2n+1)$$
so that we may identify
$$V = \ {\rm so}(2n+1) / {\rm so} (2n) \quad .$$
In this connection, we remark that the relationship between the para-fermi
statistics and Lie triple system has also been noted by
Scharfstein$^{13)}$.

Next, consider the opposite case of $V_B = 0$ with
$$\eqalignno{&V = V_F = \ <b_j , b^\dagger_j , j = 1,2,\dots, m>
&(3.5a)\cr
&<x|y>\ = -<y|x> \quad . &(3.5b)\cr}$$
In that case, we set
$$\eqalignno{&\sigma (b_j) = \sigma (b^\dagger_j) = 1 \quad , &(3.6a)\cr
&<b_j | b^\dagger_k>\ = -<b^\dagger_k | b_j>\ = \delta_{jk}
\quad , &(3.6b)\cr
&<b_j | b_k>\ =\ <b^\dagger_j | b^\dagger_k>\ = 0 \quad , &(3.6c)\cr}$$
instead of Eqs. (3.3).  Then, Eq. (2.32) is now expressed as
$$[\{x,y\},z]= 2<y|z>x + 2<x|z>y \eqno(3.7)$$
where we have rewritten $[x,y]$ here as
$$[x,y] \equiv \{ x,y\} = xy + yx \eqno(3.8)$$
in accordance with the standard convention since $(-1)^{xy} = -1$.
This gives the para-boson statistics with
$$M = \ {\rm osp} (0|2m) =\ {\rm sp} (2m) \qquad {\rm and}\qquad
L=\ {\rm osp} (1|2m) \quad .$$
Consider now the general case of
$$\eqalignno{&V= \{ a_j,a^\dagger_j,b_k,b^\dagger_k,j=1,2,\dots ,n,
k=1,2,\dots,m\}
&(3.9a)\cr
&N_B = 2n \quad , \quad N_F = 2m \quad . &(3.9b)\cr}$$
We choose the bilinear form $<x|y>$ to satisfy Eqs. (3.2) and
(3.6) as well as
$$<a_j|b_k>\ =\ <a_j |b^\dagger_k >\ = \ <a^\dagger_j | b_k>\ =
\ <a^\dagger_j | b^\dagger_k>\ = 0 \eqno(3.10)$$
for $j=1,2,\dots,n$ and $k=1,2,\dots,m$.  Again choosing $\lambda =2$, Eqs.
(2.11) and (2.32) give
$$[[x,y],z] = 2<y|z>x - 2(-1)^{xy} <x|z>y \quad . \eqno(3.11)$$
If we restrict ourselves to either the subspace Eqs. (3.2) or (3.5), then
this reproduces the results for the para-fermion and para-boson statistics.
 However, if we choose now $x=a_j,\ y=b_k,\ {\rm and}\
z = a^\dagger_\ell$ in Eq. (3.11), it gives
$$\left[ \left[ a_j , b_k \right],a^\dagger_\ell \right] =
2 \delta_{j \ell} b_k \not= 0 \quad .$$
In other words, the bosonic operator $b_k$ does not commute with the
fermionic operator $a_j$, any longer.

\medskip

\noindent {\bf 3.3 \underbar{Example 3}}

As we will show in the Appendix, this case implies the Lie-super algebra of
the type$^{3)}$ A(n$|$m) or s$\ell$(n$|$m) for $M$ and $L$;
$$\eqalignno{M &= \ {\rm s}\ell \bigg( {1 \over 2}\ N_B \bigg|
{1 \over 2}\ N_F \bigg) \oplus u(1) \quad , &(3.12a)\cr
L &= \ {\rm s}\ell \bigg( {1 \over 2}\ N_B + 1 \bigg| {1 \over 2}\
N_F \bigg) \quad , &(3.12b)\cr}$$
provided that we have $\lambda \not= 0$ with $<x|y>$ being non-degenerate.
However, we will consider only the case of $V_B = 0$ here,
$$V = V_F = \left\{ b_j , b^\dagger_j , j=1,2,\dots,m \right\}$$
as in Eq. (3.5) with Eq. (3.6).  The operations of $P$ upon $b_j$ and
$b^\dagger_j$ are assumed moveover to satisfy
$$P b_j = b_j \quad , \quad Pb^\dagger_j = -b^\dagger_j \quad , \quad
(j=1,2,\dots,n) \eqno(3.13)$$
which satisfy the conditions Eqs. (2.12) and (2.13) with $\lambda = 1$.
Eq. (2.32) with $\lambda = 1$ now becomes
$$\eqalign{[ \{ x,y \},z ]
=\ <y|Pz>Px\  &+ \ <x|Pz>Py - 2<x|Py>Pz\cr
 &+\
<y|z>x\  + \ <x|z>y \cr} \eqno(3.14)$$
where we have adopted the notation of Eq. (3.8) in order to explicitly
exhibit the symmetry $x \leftrightarrow y$.  Using Eqs. (3.13) and (3.6),
we calculate
$$\eqalignno{\big[ \big\{ b_j , b^\dagger_k \big\}, b_\ell \big] &=
2 \delta_{jk} b_\ell - 2 \delta_{k \ell} b_j
\quad , &(3.15a)\cr
\big[ \big\{ b_j , b^\dagger_k \big\}, b^\dagger_\ell \big] &=
2 \delta_{j\ell} b^\dagger_k - 2 \delta_{jk} b_\ell^\dagger
\quad , &(3.15b)\cr
\big[ \big\{ b_j , b_k \big\}, z \big] &=
\big[ \big\{
b^\dagger_j , b^\dagger_k \big\}, z \big] = 0
\quad , &(3.15c)\cr}$$
where $z$ in Eq. (3.15c) stands for $z = b_\ell$ or $b^\dagger_\ell$.  If
we renormalize $b_j$ and $b^\dagger_j$ suitably, and if we replace Eq.
 (3.15c) by the stronger relation of
$$\left\{ b_j , b_k \right\} = \left\{ b^\dagger_j ,
b^\dagger_k \right\} = 0 \quad , \eqno(3.16)$$
then this reproduces the s$\ell$(1$|$3) model of Palev$^{4),5)}$ for
$m=3$.  Actually, Eq. (3.16) is more natural than Eq. (3.15c) in our
approach by the following reason.  Returning to the original triple product
notation, Eq. (3.15c) is the rewriting of
$$\left[ b_j, b_k , z \right] = \left[
b^\dagger_j , b^\dagger_k , z \right] = 0$$
for any $z \ \epsilon\ V$.  Hence in terms of the left-multiplication
operators, this implies the validity of
$$L \left( b_j , b_k \right) = L \left( b^\dagger_j , b^\dagger_k
\right) = 0
\eqno(3.17)$$
where we used the notation of Eq. (2.22).  Since we impose the condition
Eq. (2.29a), i.e. $[x,y] = L(x,y)$ for $x,\ y\ \epsilon\ V$, this in turn
implies the validity of Eq. (3.16).  Therefore, the correct final result
should be Eqs. (3.15a), (3.15b) and (3.16).  More details on this point
will be found also in Appendix A.

Palev$^{4),5)}$ has shown that the harmonic oscillator Hamiltonian for this
model in
3-~dimensional space has a finite energy spectrum in contrast to
the standard theory.

\medskip

\noindent {\bf 3.4 \underbar{Example 4}}

As we will show in the Appendix, this case leads to osp(n$|$m) type
Lie-super algebras.  However, in order to compare our result with that of
ref. 5, we will again restrict ourselves for a consideration of the purely
fermionic space $V= V_F$ as in Eqs. (3.5) and (3.6).  Note that $b_j$ and
$b^\dagger_j$ are nevertheless bosonic annihilation and creation operators.
 We choose $\lambda = 1$ and $c = \sqrt{-1}$ in Eq. (2.16a) so that
$E_\alpha \ (\alpha = 1,2,3)$ may be identified with the Pauli matrices
$\sigma_\alpha$.  Writing
$$E_\pm = E_1 \pm \sqrt{-1} \ E_2 \quad , \eqno(3.18)$$
the actions of $E_\alpha\ (\alpha = 3,\pm)$ to $x\  \epsilon\  V$ are now
given by
$$\eqalignno{&E_3 b_j = b_j \quad ,
\quad  E_3 b^\dagger_j = -b^\dagger_j \quad , &(3.19a)\cr
&E_+ b_j = 0 \quad , \quad E_- b^\dagger_j = 0 \quad , &(3.19b)\cr
&E_- b_j = 2b_j^\dagger \quad , \quad E_+ b^\dagger_j
= 2 b_j \quad . &(3.19c)\cr}$$
Our fundamental relations is now written as
$$\eqalign{[\{x,y\},z] = &<y|E_3z>E_3x\  + \ <x|E_3z>E_3 y -
2<x|E_3y>E_3z\cr
&+ {1 \over 2}\ \{<y|E_+z>E_-x\  + \ <x|E_+z>E_-y - 2<x|E_+y>E_-z\cr
&+ \ <y|E_-z>E_+x\  + \ <x|E_-z>E_+y - 2<x|E_-y>E_+z\}\cr
&+\ <y|z>x\  + \ <x|z>y \quad . \cr}\eqno(3.20)$$
\noindent From this, we calculate
$$\eqalignno{\big[\big\{ b_j,b^\dagger_k \big\}, b_\ell \big] &=
2 \big\{ -\delta_{k \ell}b_j + \delta_{jk} b_\ell
+ \delta_{j \ell} b_k \big\} &(3.21a)\cr
\big[\big\{ b_j,b^\dagger_k \big\}, b_\ell^\dagger \big] &=
2 \big\{ -\delta_{k \ell}b_j^\dagger - \delta_{jk} b_\ell^\dagger
+ \delta_{j \ell} b_k^\dagger \big\} &(3.21b)\cr
\big[\big\{ b_j,b_k \big\}, b_\ell^\dagger \big] &=
-4 \delta_{jk}b_\ell &(3.21c)\cr
\big[\big\{ b_j^\dagger,b^\dagger_k \big\}, b_\ell \big] &=
 4 \delta_{jk}b^\dagger_\ell &(3.21d)\cr
\big[\big\{ b_j,b_k \big\}, b_\ell \big] &=
\big[ \big\{  b_j^\dagger , b_k^\dagger \big\},
 b_\ell^\dagger \big] = 0 \quad . &(3.21e)\cr}$$
Moreover, we will show in the Appendix the validity of
$$\eqalignno{&L\big( b_j , b_k \big) = -2 \delta_{jk} E_+ &(3.22a)\cr
&L\big( b^\dagger_j , b^\dagger_k \big) = 2 \delta_{jk} E_- &(3.22b)\cr
&L \big( b_j , b^\dagger_k \big) + L \big( b_k , b^\dagger_j \big) =
4 \delta_{jk} E_3 &(3.22c)\cr}$$
as in Eqs. (A.46), where $E_+,\ E_-\ {\rm and}\ E_3$ generate a su(2) Lie
algebra (see Eqs. (A.38)).  Then, as has been discussed in the previous
case of 3.3, we replace Eqs. (3.22) by the operator equations
$$\eqalignno{&\big\{ b_j , b_k \big\} = - 2 \delta_{jk} E_+ &(3.23a)\cr
&\big\{ b^\dagger_j , b^\dagger_k \big\} = 2 \delta_{jk} E_- &(3.23b)\cr
&\big\{ b_j , b^\dagger_k \big\} + \big\{ b_k , b_j^\dagger \big\} =
4 \delta_{jk} E_3 \quad . &(3.23c)\cr}$$
Then, the relations Eqs. (3.21a), (3.21b), and (3.21e) together with Eqs.
(3.23) essentially reproduce the commutation relations proposed in ref. 5
for the 3-dimensional model of $L =$ osp(3$|$2) for $m=3$, if we normalize
$b_j$ and $b^\dagger_j$ by $b_j = \sqrt{3}\ a^-_j$ and $b^\dagger_j =
\sqrt{3}\ a^+_j$. As has been noted by Palev and Stoilova, this model
has many interesting features.  For example, the spin 1/2 realization is
possible now although we work only with bosonic operators.

\vfill\eject

\medskip

\noindent {\bf 3.5 \underbar{Other Examples}}

The example 5 given by Eqs. (2.19) and (2.20) is really a generalization of
the example 2.  Indeed, if we choose $n=1$, then Eq. (2.20) will reduce to
Eq. (2.11).  For the general value of $n$, the corresponding Lie-super
algebras are again of the type osp(n$|$m), but we will not go into its
detail.  Here, we will discuss the singluar case of $\lambda =0$ in Eqs.
(2.13) and (2.14) so that we have $P^2 =0$, identically.  For simplicity, we
suppose
$$V = V_B = \ <a_j , a^\dagger_j , c_j, c^\dagger_j, j=1,2,\dots,n>
\eqno(3.24)$$
with
$$\eqalignno{P a_j &= c_j \quad , \quad Pc_j = 0 \quad , &(3.25a)\cr
P a_j^\dagger &= c^\dagger_j \quad , \quad Pc_j^\dagger
 = 0 \quad . &(3.25b)
\cr}$$
Moreover, we may assume
$$<a_j | a^\dagger_k>\ = \gamma_{jk} \eqno(3.26a)$$
and
$$<c_j | a^\dagger_k>\ = -<a_j | c^\dagger_k>\ = \beta_{jk}
\eqno(3.26b)$$
for some arbitrary constants $\gamma_{jk}$ and $\beta_{jk}$, while all
other  products (except for those obtained from $<y|x>\ =\ <x|y>$), such
as $<c_j |a_k>$ and $<c_j | c^\dagger_k>$, are zero.  Then, only non-zero
triple products, apart from those obtained from $[x,y,z] = - [y,x,z]$, are
calculated to be
\item{(i)} $\left[ a_j, a_k, a^\dagger_\ell \right] = -
\beta_{k \ell} c_j + \beta_{j \ell} c_k \quad ,$ \hfill (3.27a)

\item{(ii)} $\left[ a_j^\dagger, a_k^\dagger, a_\ell \right] =
\beta_{\ell k} c_j^\dagger - \beta_{\ell j} c_k^\dagger
 \quad ,$ \hfill (3.27b)

\item{(iii)} $\left[ a_j, a_k^\dagger, a_\ell \right] =
\beta_{\ell k} c_j + 2 \beta_{jk} c_\ell \quad ,$ \hfill (3.27c)

\item{(iv)} $\left[ a_j, a_k^\dagger, a^\dagger_\ell \right] =
\beta_{j \ell} c_k^\dagger + 2 \beta_{jk} c_\ell^\dagger \quad .$
 \hfill (3.27d)

\noindent Especially, we have
$$L \left( c_j , x \right) = L\left( c^\dagger_j , x \right) = 0
\eqno(3.28)$$
for any $x = a_k,\ a^\dagger_k,\ c_k,\ {\rm and}\
c^\dagger_k$.  Then, it is easy to see
$$\left[ L_{u,v},L_{x,y}\right] = 0 \eqno(3.29)$$
so that the Lie algebra $M$ is purely Abelian, while $L$ is nilpotent.
Finally, we simply remark that the condition Eq. (3.29) can be used to
construct solutions of both Yang-Baxter and classical Yang-Baxter
equations$^{11),14)}$.

\medskip

\noindent {\bf \underbar{Acknowledgment}}

This paper is supported in part by the U.S. Department of Energy Grant.
No.

\noindent DE-FG-02-91ER40685.

\vfill\eject

\noindent {\bf Appendix A. \underbar{Construction of Lie-super Algebras}}

\noindent {\bf A.1 \underbar{The Case of Example 2}}

Consider first the example 2 so that we have
$$[x,y,z] =\ <y|z>x - (-1)^{yz} <x |z>y \eqno(A.1)$$
where we have set $\lambda =1$ in Eq. (2.11) for simplicity.  Let
 $e_1,\ e_2\ \dots,\ e_N$ be a basis of $V$ and set
$$\eqalignno{\sigma_j &= \sigma (e_j) \quad , &(A.2a)\cr
g_{jk} &= \ <e_j | e_k> \quad . &(A.2b)\cr}$$
Then, Eqs. (2.8) imply the validity of
$$\eqalignno{g_{kj} &= (-1)^{\sigma_j \sigma_k} g_{jk} &(A.3a)\cr
g_{jk} &= 0 \quad , \quad {\rm if}\quad
\sigma_j \not= \sigma_k &(A.3b)\cr}$$
and Eq. (A.1) becomes
$$\left[ e_j , e_k , e_\ell \right] = g_{k \ell} e_j
- (-1)^{\sigma_k \sigma_\ell} g_{j \ell} e_k \quad . \eqno(A.4)$$
Setting further
$$L_{j,k} = L \left( e_j , e_k \right) \quad , \eqno(A.5)$$
Eq. (2.24b) gives
$$\eqalign{\big[ L_{j,k}, L_{\ell ,m} \big] = \
&g_{k \ell} L_{j,m} - (-1)^{\sigma_k \sigma_\ell}
g_{j \ell} L_{k,m}\cr
& + (-1)^{(\sigma_j + \sigma_k)\sigma_\ell}
\big\{ g_{km} L_{\ell, j} - (-1)^{\sigma_k \sigma_m}
g_{jm} L_{\ell, k} \big\} \quad , \cr}\eqno(A.6)$$
so that we find the ortho-symplectic Lie-super algebra
$$M = \ {\rm osp} \left( N_B | N_F \right) \eqno(A.7)$$
for Eq. (A.6).  Similarly, Eqs. (2.29) lead to
$$\eqalignno{&\big[ e_j , e_k \big] = L_{j,k} &(A.8a)\cr
&\big[ L_{j,k}, e_\ell \big] = g_{k \ell} e_j -
(-1)^{\sigma_k \sigma_\ell} g_{j \ell} e_k \quad . &(A.8b)\cr}$$
For simplicity, set
$$B = N + 1 \eqno(A.9)$$
and identify
$$e_j = L_{j,B} = - L_{B, j} \eqno(A.10a)$$
with the assignment of $\sigma_B = 0$ so that
$$\sigma \left( L_{j,B} \right) = \left( \sigma_j +
\sigma_B \right) \ ({\rm mod}\ 2) =
 \sigma_j \quad . \eqno(A.10b)$$
Further, we introduce $g_{j,B},\ g_{B,j},\ {\rm and}\ g_{B,B}$ by
$$\eqalignno{g_{j,B} &= g_{B,j} = 0 &(A.11a)\cr
g_{B,B} &= -1 \quad . &(A.11b)\cr}$$
Then, Eqs. (A.8a) and (A.8b) can be incorporated into a single equation of
Eq. (A.6) where $N_B$ is now extended into $N_B + 1$.  This implies
$$L = \ {\rm osp} \left( N_B + 1 | N_F \right) \quad . \eqno(A.12)$$
Therefore, Eq. (2.31) may be rewritten as
$$V = \ {\rm OSP} \left( N_B +1|N_F\right) / {\rm OSP} \left( N_B | N_F
\right) \quad . \eqno(A.13)$$
Also, for the present system, the equality in Eq. (2.33) holds valid,
indicating that there is no extra linear relations among $L_{j,k}$'s for
this case.

\medskip
\vfill\eject

\noindent {\bf A.2 \underbar{Example 3}}

We choose $\lambda =1$ again in Eqs. (2.13) and (2.14) so that
$$\eqalign{[x,y,z] = Px &<y|Pz> \ -\  (-1)^{xy} Py <x|Pz>\cr
&- 2 Pz <x|Py> \ + \ <y|z>x - (-1)^{xy} <x|z>y \quad . \cr}\eqno(A.14)$$
Since $P^2 = Id$, we can classify the basis of $V$ according to
eigenvectors of $P$.  For simplicity, we assume here that $P$ possesses
equal numbers of eigenvectors for $P=1$ and $P=-1$ with the same gradings,
i.e.
$$Pe_j = e_j \quad , \quad P \overline e_j = - \overline e_j
\quad , \quad (j = 1,2,\dots,n) \eqno(A.15a)$$
$$\sigma (e_j) = \sigma (\overline e_j )= \sigma_j \quad , \quad (j =1,2,
\dots,n) \quad , \eqno(A.15b)$$
and $V$ can be expressed as
$$V = \ <e_1 , e_2, \dots , e_n, \overline e_1 , \overline e_2 , \dots ,
\overline e_n> \eqno(A.16a)$$
$$N_B = 2n_B \quad , \quad N_F = 2n_F \quad , \quad
n = n_B +n_F \quad . \eqno(A.16b)$$
We then calculate
$$\eqalignno{&<e_j |e_k>\ = \ <\overline e_j | \overline e_k>\ = 0 &(A.17a)
\cr
&g_{j \overline k} = \ <e_j | \overline e_k>\ = (-1)^{\sigma_j \sigma_k}
< \overline e_k | e_j>\ = (-1)^{\sigma_j \sigma_k}
g_{\overline k j} &(A.17b)\cr}$$
since Eq. (2.13b) for example leads to
$$<e_j|e_k>\ =\ <Pe_j |e_k>\ = -<e_j |Pe_k>\ = -<e_j|e_k>\ =0
\quad .$$
Moreover, we find
$$\eqalignno{\big[ e_j,e_k,z\big] &= \big[ \overline e_j , \overline
e_k , z \big] = 0 \quad , \quad \big( z= e_\ell
\ {\rm or}\ \overline e_\ell \big) \quad , &(A.18a)\cr
\big[ e_j, \overline e_k, e_\ell \big] &= 2 \big\{ g_{j \overline k}
e_\ell + (-1)^{\sigma_k \sigma_\ell}
g_{\ell \overline k} e_j\big\} \quad ,
&(A.18b)\cr
\big[ e_j, \overline e_k, \overline e_\ell \big]
 &= -2 \big\{ g_{j \overline k}
\overline e_\ell + (-1)^{\sigma_k \sigma_\ell}
g_{j \overline \ell} \overline e_k \big\} \quad ,
&(A.18c)\cr}$$
Hence, setting
$$\eqalignno{L_{j,k} &= L \big( e_j , e_k \big) &(A.19a)\cr
L_{\overline j, \overline k} &= L \big( \overline e_j , \overline e_k
\big) &(A.19b)\cr
L_{j, \overline k} &= L \big(e_j , \overline e_k
\big) = -(-1)^{\sigma_j \sigma_k} L
\big( \overline e_k , e_j \big)
\quad , &(A.19c)\cr}$$
Eq. (A.18a) implies
$$L_{j,k} = L_{\overline j , \overline k} = 0
\eqno(A.20)$$
while Eq. (2.24) gives
$$\left[ L_{j, \overline k}, L_{\ell , \overline m}\right] =
2 \left\{ g_{\overline k \ell} L_{j, \overline m} -
(-1)^{(\sigma_j + \sigma_k) \sigma_\ell + \sigma_k \sigma_m}
g_{j, \overline m} L_{\ell , \overline k} \right\} \quad .$$
Therefore, setting
$$X^k_j = {1 \over 2}\ L_{j, \overline k} - g_{j, \overline k}
P \eqno(A.21)$$
and noting $g_{j, \overline k} =0$ unless $\sigma_j = \sigma_k$, we find
$$\eqalignno{\big[ X^k_j , X^m_\ell \big] &= g_{\overline k \ell}
X^m_j - (-1)^{(\sigma_j + \sigma_k)(\sigma_\ell + \sigma_m )}
g_{\overline m j} X^k_\ell &(A.22a)\cr
\big[ X^k_j , P \big] &= 0 &(A.22b)\cr}$$
which defines the Lie-super algebra of the type
$$M = u(1) \oplus \ {\rm s}\ell \left( {1 \over 2}\ N_B \bigg|
{1 \over 2}\ N_F \right) \quad . \eqno(A.23)$$
Note that both $N_B = 2n_B$ and $N_F = 2n_F$ are even.

Next, we will determine the Lie-super algebra of $L$.  We first find
$$\eqalignno{&\sum^n_{j,k=1} g^{j \overline k} X^k_j = P
 \quad , &(A.24a)\cr
&\big[e_j , \overline e_k \big] = L_{j \overline k} = 2
\big( X^k_j + g_{j \overline k} P \big) \quad , &(A.24b)\cr
&\big[ X^k_j , e_\ell \big] = g_{\overline k \ell} e_j
 \quad , &(A.24c)\cr
&\big[ X^k_j , \overline e_\ell \big] =
-(-1)^{\sigma_k \sigma_\ell} g_{j \overline \ell} \overline e_k
\quad , &(A.24d)\cr
&\big[ P , e_\ell \big] = e_\ell \quad ,  &(A.24e)\cr
&\big[ P, \overline e_\ell \big] = - \overline e_\ell
 \quad , &(A.24f)\cr
&[P,P] = \big[ e_j , e_k \big] = \big[ \overline e_j , \overline e_k \big]
=0 \quad . &(A.24g)\cr}$$
Setting for simplicity
$$B = n + 1 \quad , \quad \sigma_B = 0 \eqno(A.25a)$$
and introducing
$$\eqalignno{&g_{B, \overline B} = 1 \quad , \quad g_{B, \overline j}
= g_{j, \overline B} = 0 \quad , &(A.25b)\cr
&X^B_B = -P \quad , \quad X^B_j = {1 \over \sqrt{2}}\ e_j \quad ,
\quad X^j_B = {1 \over \sqrt{2}}\ \overline e_j \quad , &(A.25c)\cr}$$
we see that Eqs. (A.24) can be incorporated into Eq. (A.22a) by formally
extending $n_B$ into $n_B \rightarrow n_B + 1$.  With
$\sum^{n+1}_{j,k =1} g^{j \overline k} X^k_j = 0$, we find then
$$L= \ {\rm s}\ell \left( {1 \over 2}\ N_B + 1 \bigg| {1 \over 2}\ N_F \right)
\quad . \eqno(A.26)$$

Finally, the case of $\lambda = 0$ in Eqs. (2.13) and (2.14) is singular,
but we will not go into its analysis here.

\medskip

\noindent {\bf A.3 \underbar{Example 4}}

Again, we choose $\lambda = 1$ and $c = \sqrt{-1}$ in Eq. (2.16a) so that
$E_\alpha (\alpha = 1,2,3)$ behave as Pauli's $2 \times 2$ matrices.
Introducing
$$E_\pm = E_1 \pm \sqrt{-1}\ E_2 \quad , \eqno(A.27)$$
we choose the basis of $V$ to be the same as Eq. (A.16) for simplicity.
Moreover Eq. (A.15a) is replaced now by
$$\eqalignno{&E_3 e_j = e_j \quad , \quad E_3 \overline e_j =
- \overline e_j &(A.28a)\cr
&E_+ e_j = 0 \quad , \quad E_+ \overline e_j = 2 e_j &(A.28b)\cr
&E_- e_j = 2 \overline e_j \quad , \quad E_- \overline e_j
 = 0 &(A.28c)\cr}$$
so that Eqs. (2.16) are satisfied.  For the present case, we have
$$\eqalignno{&g_{jk} = \ <e_j | e_k >\ = 0 &(A.29a)\cr
&g_{\overline j\ \overline k} = \ <\overline e_j | \overline e_k>\ = 0
&(A.28b)\cr
&g_{j \overline k} =\ <e_j | \overline e_k>\ = - (-1)^{\sigma_j \sigma_k}
g_{k \overline j} &(A.29c)\cr}$$
since we calculate
$$\eqalign{<e_j |e_k>\ &=\ <E_3 e_j | e_k>\ = -<e_j | E_3 e_k>\
= - <e_j |e_k>\ = 0 \quad ,\cr
< \overline e_j | \overline e_k>\ &=
 - <E_3 \overline e_j | \overline e_k>\ = \ <\overline e_j | E_3
\overline e_k>\
= - < \overline e_j | \overline e_k>\ = 0 \quad ,\cr
<e_j |\overline e_k>\ &= {1 \over 2} <e_j |E_- e_k>\ = -
{1 \over 2} <E_- e_j | e_k>\cr
&= - < \overline e_j |e_k>\ = - (-1)^{\sigma_j \sigma_k}
<e_k | \overline e_j> \quad . \cr}$$
It is convenient to rewrite $g_{j \overline k}$ as $f_{jk}$ i.e.,
$$\eqalignno{f_{jk} &= g_{j \overline k} = -(-1)^{\sigma_j \sigma_k}
f_{kj} &(A.30a)\cr
f_{jk} &= 0 \quad {\rm unless}\quad \sigma_j = \sigma_k \quad .
&(A.30b)\cr}$$
Then, we calculate
$$\eqalignno{[e_j, e_k, e_\ell] &= [\overline e_j , \overline e_k
, \overline e_\ell ] = 0 \quad , &(A.31a)\cr
[e_j , e_k , \overline e_\ell ]&= -4 f_{jk} e_\ell \quad ,
&(A.31b)\cr
[\overline e_j , \overline e_k , e_\ell ] &= 4
f_{jk} \overline e_\ell \quad , &(A.31c)\cr
[e_j , \overline e_k , e_\ell ] &=
-(-1)^{\sigma_j \sigma_k} [ \overline e_k , e_j , e_\ell ] \cr
&= 2 \big\{ -f_{k \ell} e_j + f_{jk} e_\ell -
(-1)^{\sigma_k \sigma_\ell} f_{j \ell} e_k \big\} \quad ,
&(A.31d)\cr
[e_j , \overline e_k , \overline e_\ell ] &=
-(-1)^{\sigma_j \sigma_k} [ \overline e_k , e_j , \overline e_\ell ] \cr
&= - 2 \big\{ f_{k \ell} \overline e_j + f_{jk} \overline e_\ell +
(-1)^{\sigma_k \sigma_\ell} f_{j \ell} \overline e_k \big\} \quad .
&(A.31e)\cr}$$
We defined $L_{j,k}$ etc. again by Eqs. (A.19).  However, we have many
linearly dependent relations among these for the present case.
For example, if we calculate $[L_{j,k}, L_{\ell , \overline m}]$ in Eq.
(2.34) we find a constraint equation
$$\eqalign{4 f_{\ell , m} L_{j,k} &- 4 f_{j,k} L_{\ell , m} - 2
f_{m,j} L_{\ell , k} - 2(-1)^{\sigma_m \sigma_j} f_{\ell , j}
L_{m,k}\cr
&- 2(-1)^{(\sigma_\ell + \sigma_m)\sigma_j}
\big\{f_{m,k} L_{j , \ell} + (-1)^{\sigma_m \sigma_k} f_{\ell , k}
L_{j,m} \big\} = 0 \quad . \cr}$$
However, these constraints are simplified as follows. Comparing Eqs. (A.31)
with Eqs. (A.28), it is not hard to see the validity of
$$\eqalignno{L_{j,k} &= -2 f_{j,k} E_+ &(A.32a)\cr
L_{\overline j,\ \overline k} &= 2 f_{j,k} E_- &(A.32b)\cr}$$
while we have an additional relation
$$\eqalignno{&L_{j, \overline k} - (-1)^{\sigma_j \sigma_k} L_{k, \overline j}
= 4 f_{jk} E_3 &(A.33a)\cr
&\big[ L_{j,\overline k}, E_3 \big] = 0 \quad . &(A.33b)\cr}$$
Introducing $Y_{jk}$ by
$$Y_{jk} = {1 \over 2}\ L_{j, \overline k} - f_{jk}
E_3 = {1 \over 4}\ \left\{ L_{j, \overline k}
 + (-1)^{\sigma_j \sigma_k}
L_{k, \overline j} \right\} \quad , \eqno(A.34)$$
we can conversely express
$$L_{j, \overline k} = -(-1)^{\sigma_j \sigma_k}
L_{\overline k,j} = 2 Y_{j,k}
+ 2 f_{jk} E_3 \quad . \eqno(A.35)$$
We can easily verify now that $Y_{j,k}$'s satisfy
$$\eqalignno{&Y_{j,k} = (-1)^{\sigma_j \sigma_k} Y_{k,j} \quad ,
&(A.36a)\cr
&\big[ Y_{j,k} , Y_{\ell ,m} \big] = - f_{k, \ell} Y_{j,m}
- (-1)^{\sigma_k \sigma_\ell} f_{j, \ell} Y_{k,m}\cr
&\qquad\qquad\qquad\qquad - (-1)^{(\sigma_j + \sigma_k)\sigma_\ell} \big\{
f_{k,m} Y_{\ell j} + (-1)^{\sigma_k \sigma_m}
f_{jm} Y_{\ell k} \big\} &(A.36b) \cr}$$
so that it defines the Lie-super algebra osp$({1 \over 2}\ N_F | {1
\over 2}\ N_B )$.  Note that the order of $N_B$ and $N_F$ is now
interchanged in comparison to Eq. (A.23).  Moreover, we have
$$\left[ Y_{j,k} , E_+ \right] = \left[ Y_{j,k} , E_- \right] = \left[
Y_{j,k} , E_3 \right] = 0 \quad . \eqno(A.37)$$
Noting further that
$$\eqalignno{\big[ E_+ , E_3 \big] &= -2E_+ &(A.38a)\cr
\big[ E_- , E_3 \big] &= 2 E_- &(A.38b)\cr
\big[ E_+ , E_- \big] &= 4 E_3 &(A.38c)\cr}$$
is the su(2) Lie algebra, we conclude
$$M =\  {\rm su}(2) \oplus\  {\rm osp} \left( {1 \over 2}
\ N_F \bigg| {1 \over 2}\ N_B \right) \quad . \eqno(A.39)$$
Now, we will turn our attention to the Lie-super algebra $L$.  We
calculate from Eqs. (2.29)
$$\eqalignno{\big[ Y_{j,k} , e_\ell \big] &= -f_{k \ell} e_j
- (-1)^{\sigma_k \sigma_\ell} f_{j \ell} e_k \quad , &(A.40a)\cr
\big[ Y_{j,k} , \overline e_\ell \big] &= -f_{k \ell} \overline e_j
- (-1)^{\sigma_k \sigma_\ell} f_{j \ell} \overline e_k \quad , &(A.40b)\cr
\big[ e_j , e_k \big] &= L_{j,k} = -2 f_{jk}
E_+ &(A.40c)\cr
\big[ \overline  e_j , \overline e_k \big] &=
 L_{ \overline j,\   \overline k} = 2 f_{jk}
E_- &(A.40d)\cr
\big[   e_j , \overline e_k \big] &=
 L_{ j, \overline k} = 2 Y_{jk} + 2 f_{jk}
E_3 &(A.40e)\cr}$$
as well as
$$\eqalignno{\big[ E_+ , e_j \big] &= \big[ E_- , \overline e_j \big]
 = 0 \quad ,  &(A.41a)\cr
\big[ E_+ , \overline e_j \big] &= 2 \big[ E_3 ,  e_j \big]
 = 2 e_j \quad , &(A.41b)\cr
\big[ E_- , e_j \big] &= -2 \big[ E_3 , \overline e_j \big]
 = 2 \overline e_j \quad . &(A.41c)\cr}$$
Setting further $B = n + 1$ with $\sigma_B = \sigma_{B+1} = 0$ as well as
$$\eqalignno{&e_j = \sqrt{2} \ Y_{j,B} \quad , \quad \overline e_j
= \sqrt{2}\ Y_{j,B+1}  &(A.42a)\cr
&f_{j,B} = f_{j,B+1} = f_{B,j} =
f_{B+1,j} = 0 &(A.42b)\cr
&f_{B,B+1} = -f_{B+1,B} = -1 &(A.42c)\cr
&f_{B,B} = f_{B+1,B+1} = 0 &(A.42d)\cr}$$
together with
$$\eqalignno{E_+ &= Y_{B,B} &(A.43a)\cr
E_- &= - Y_{B+1,B+1} &(A.43b)\cr
E_3 &= -Y_{B,B+1} = - Y_{B+1,B}\quad , &(A.43c)\cr}$$
these relations are again incorporated into a single equation of Eqs.
 (A.36), if we extend $n_B$ into $n_B + 2$.
 Therefore, we find
$$L = \ {\rm osp} \left( {1 \over 2}\ N_F \bigg| {1 \over 2}\
N_B + 2 \right)\quad . \eqno(A.44)$$

When we have $N_B = 0,$ and $N_F = 2m$ as in section (3.4), these give
$$\eqalignno{M &= \ {\rm su} (2) \oplus  {\rm so}(m) &(A.45a)\cr
L &= \ {\rm osp}(m|2) &(A.45b)\cr}$$
with
$$\eqalignno{&f_{jk} = \delta_{jk} \quad {\rm for}
\quad j,k = 1,2,\dots, m \quad , &(A.46a)\cr
&L_{j,k} = -2 \delta_{jk} E_+ &(A.46b)\cr
&L_{\overline j,\ \overline k} = 2 \delta_{jk} E_- &(A.46c)\cr
&L_{j, \overline k} + L_{k, \overline j} = 4 \delta_{jk} E_3 &(A.46d)\cr}$$
since we have $\sigma_j = \sigma_k = 1$.

\medskip

\noindent {\bf Appendix B.
\underbar{Relationship with Ortho-Symplectic Super-Triple System}}

Here, we will discuss a connection between ortho-symplectic super triple
systems and Lie-super triple systems.

Let $V$ be a super-space as in section 2.  Suppose that a triple product
$x\ y\ z\  :\ V \otimes V \otimes V \rightarrow V$ satisfy the following
ans\"atz:

\item{(i)} $\sigma (xyz) = \{ \sigma (x) + \sigma (y) +
\sigma (z) \}$ (mod 2) \hfill (A.47a)

\item{(ii)} $x\ y\ z + (-1)^{xy} y\ x\ z = 0$ \hfill (A.47b)

\item{(iii)} $x\ y\ z + (-1)^{yz} x\ z\ y= 2 \lambda <y|z>x\ -
\lambda <x|y>z\ - \lambda (-1)^{yz} <x|z>y$ \hfill (A.47c)

\item{(iv)} $u\ v (xyz) = (uvx)yz + (-1)^{(u+v)x} x(uvy)z
+ (-1)^{(u+v)(x+y)} x\ y(uvz)$ \hfill (A.47d)

\item{(v)} $<uvx|y>\ = -(-1)^{(u+v)x} <x|uvy>$ \hfill (A.47e)

\noindent for a constant $\lambda$.  Then, $V$ is called a ortho-symplectic
super-triple system (hereafter referred to as OSST), which generalizes the
notion of both orthogonal and symplectic triple systems discussed
elsewhere$^{15)}$.  We also note that Eq. (A.47e) is a consequence of other
postulates Eqs. (A.47a)--(A.47d), if $\lambda \not= 0$.

Let $e_1,\ e_2, \ \dots,\ e_N\ (N = \ {\rm Dim}\ V)$ be a basis of $V$ and
set
$$g_{jk} =\ <e_j |e_k>\ = (-1)^{\sigma_j \sigma_k} g_{kj}
\eqno(A.48)$$
as before with its inverse $g^{jk}$.  Setting
$$e^j = \sum^N_{k=1} g^{jk} e_k \quad , \quad \sigma (e^j) =
\sigma (e_j) \equiv \sigma_j \quad , \eqno(A.49)$$
we see then
$$\eqalignno{&<e^j | e_k>\ = \delta^j_k \quad , &(A.50a)\cr
&x = \sum^N_{j=1} <x|e_j>e^j =
\sum^N_{j=1} e_j <e^j |x> \quad . &(A.50b)\cr}$$
Given a OSST product $x\ y\ z$, then we set
$$C[x,y,z] = \sum^N_{j=1} \left( xye_j
\right) e^jz + {1 \over 3}\ \lambda \left( N_0 - 16 \right) x \ y\ z
\eqno(A.51)$$
for a constant $C$, where $N_0$ is given by
$$N_0 = N_B - N_F \quad . \eqno(A.52)$$
Then, a simple generalization of the result of ref. 15 shows that
$[x,y,z]$ is a Lie-super triple system, provided that $C \not= 0$.

Let us now return to some examples.  First, Eq. (2.11) given in the example
2 is also a OSST with identification of $x\ y\ z = [x,y,z]$ with
$C = {1 \over 3}\ \lambda \left( N_0 - 10 \right)$.  Next, let $P$ be as in
the example 3 satisfying conditions Eqs. (2.12) and (2.13) and set
$$\eqalign{x\ y\ z = Px<y|Pz>\ &- (-1)^{xy} Py <x|Pz>\ + \ Pz<x|Py>\cr
&+ \lambda \{ <y|z>x\ - (-1)^{xy} <x|z>y\} \quad . \cr} \eqno(A.53)$$
Then, we can show that this defines a OSST.  Moreover, Eq. (A.51)
reproduces Eq. (2.14) with
$$C = {1 \over 3}\ \lambda \left( N_0 -4 \right)\quad . \eqno(A.54)$$
Finally, for $E_\alpha (\alpha = 1,2,3)$ given by Eqs. (2.15) and (2.16),
we define
$$\eqalign{x\ y\ z = \sum^3_{\alpha = 1} \big\{ &E_\alpha x < y | E_\alpha
z>  -\  (-1)^{xy} E_\alpha y <x |E_\alpha z> \ +\ E_\alpha
z<x|E_\alpha y>\big\}\cr
&+ \lambda \big\{ <y | z>x\ -\ (-1)^{xy} <x|z>y\big\}\cr}\eqno(A.55)$$
which can be shown to satisfy all axioms of a OSST.  Moreover, Eq.
(A.51) leads to the result of Eq. (2.17) with
$$C = {1 \over 3}\ \lambda \left( N_0 + 8 \right) \quad . \eqno(A.56)$$
There are, however, many examples that we have $C=0$ identically in Eq.
(A.51).  Such cases are useful to construct solutions of Yang-Baxter
equations as we have noted elsewhere$^{8),15)}$.

\vfill\eject

\noindent {\bf \underbar{References}}

\item{1.} Y. Ohnuki and S. Kamefuchi, \underbar{Quantum Field Theory and
 Parastatistics}, (University of Tokyo Press, Springer-Verlag, Berlin,
1982) and earlier references quoted therein.

\item{2.} M. Omate, Y. Ohnuki and S. Kamefuchi, Prog. Theor. Phys.
{\bf 56}, 1948 (1976).

\item{3.} M. Scheunert, \underbar{The Theory of Lie Superalgebras},
(Springer-Verlag, Berlin-Heidelberg-New York, 1979), V. G. Kac, Adv. Math.
{\bf 26}, 8 (1977).

\item{4.} T. D. Palev, Jour. Math. Phys. {\bf 26}, 1640 (1985);
{\bf 27}, 1994 (1986); {\bf 28}, 272 (1987); {\bf 28}, 2280 (1987);
{\bf 29}, 2589 (1988).

\item{5.} T. D. Palev and N. I. Stoilova, ICTP preprint IC/93/190 (also
HEP-TH/9307102) unpublished (1993).

\item{6.} W. G. Lister, Am. J. Math. {\bf 89}, 787 (1952).

\item{7.} K. Yamaguchi, J. Sci. Hiroshima University {\bf A21}, 155 (1958).

\item{8.} S. Okubo, University of Rochester Report UR-1312 (1993)  to
appear in the Proceedings of the 15th MRST Meeting held at Syracuse
University on May 14-15, 1993.

\item{9.} K. Yamaguchi, in S\=urikagaku K\=oky\=uroku
{\bf 308}, University of Kyoto, Inst. Math. Analysis (1977) in Japanese.

\item{10.} N. Kamiya, Jour. Algebra {\bf 110}, 108 (1987).

\item{11.} S. Okubo, University of Rochester Report, UR-1319 (1993).
\item{12.} S. Helgason, \underbar{Differential Geometry of Symmetric
Spaces}, (Academic Press, New York, 1962).

\item{13.} H. Scharfstein, Preprint, P.O. Box 641, Trumbull, CT 06611,
(1985) (unpublished).

\item{14.} S. Okubo, University of Rochester Report,
UR-1335 (1994).

\item{15.} S. Okubo, Jour. Math. Phys. {\bf 34}, 3273, 3292 (1993).

\end